# Revealing the Nanostructure of Mesoporous Fuel Cell Catalyst Supports for Durable, High-Power Performance

Matthew Ko[1], Elliot Padgett[1,*], Venkata Yarlagadda[2], Anusorn Kongkanand[2], David A. Muller[1,3]

[1] School of Applied and Engineering Physics, Cornell University, Ithaca, New York 14853, United States

[2] Fuel Cell Business, Global Propulsion Systems, General Motors, Pontiac, MI 48340, United States

[3] Kavli Institute at Cornell for Nanoscale Science, Ithaca, New York 14853, United States

* current address: National Renewable Energy Laboratory, Golden, Colorado 80401-3305, United States

## Abstract

Achieving high power performance and durability with low Pt loadings are critical challenges for proton exchange membrane fuel cells. PtCo catalysts developed on new carbon black supports show promise by simultaneously providing good oxygen reduction kinetics and local oxygen transport. We investigate the role of nanoscale morphology in the performance of these catalysts supported on accessible (HSC-e and HSC-f) and conventional (Ketjen Black) porous carbons using 3D electron tomography, nitrogen sorption, and electrochemical performance measurements. We find that the accessible porous carbons have hollow interiors with mesopores that are larger and more numerous than conventional porous carbons. However, mesopore-sized openings (>2nm width) are too rare to account for significant oxygen transport. Instead we propose the primary oxygen transport pathway into the interior is through 1-2nm microporous channels permeating the carbon. The increased mesoporosity in the accessible porous carbons results in a shorter diffusion pathlength through constrictive, tortuous micropores in the support shell leading to lower local oxygen transport resistance. In durability testing, the accessible porous carbons show faster rates of electrochemical surface area loss, likely from fewer constrictive pores that would mitigate coarsening, but maintain superior high current density performance at end of life from the improved local oxygen transport.

## Introduction

Hydrogen-based proton exchange membrane fuel cells (PEMFCs) are a promising clean alternative to the combustion engine. However, their widespread adoption for vehicle applications is limited by the high cost and platinum requirements of current fuel cell technology[1,2]. Because of their high performance and relative stability as oxygen reduction catalysts, Pt and Pt-alloy nanoparticles deposited on high surface area carbon blacks have been used in commercially-available fuel cell vehicles[3] and are likely to see continued use into the future. As such, developing Pt-based catalysts to optimize performance and durability while minimizing Pt usage is essential to the success of automotive fuel cells.

The structure and morphology of the carbon support, as well as the location of the metal catalyst particles on the support, have important impacts on fuel cell performance and durabiltiy[4–6]. Particles supported on solid carbons (such as Vulcan carbon) are exposed to ionomer, leading to adsorption of ionomer on the Pt surface, which reduces oxygen reduction reaction (ORR) activity by a factor of 2-4.[6–9] On porous carbon supports (such as KetjenBlack), a large fraction of metal catalyst particles reside inside the carbon particles[10,11]. These internal particles can be shielded from ionomer if the openings of the carbon pores are smaller than the hydrodynamic size of the ionomer (~2-7 nm) when mixed in an electrode ink. Porous carbon supports can also contribute to catalyst durability by protecting against catalyst coarsening by particle migration and coalescence.[5] While internal catalyst particles have higher ORR activity and do not have significant proton transport challenges,[12,13] they can experience oxygen transport problems from constrictive pores. This is particularly problematic for high power operation, where a high reactant flux must reach the active catalyst surface.

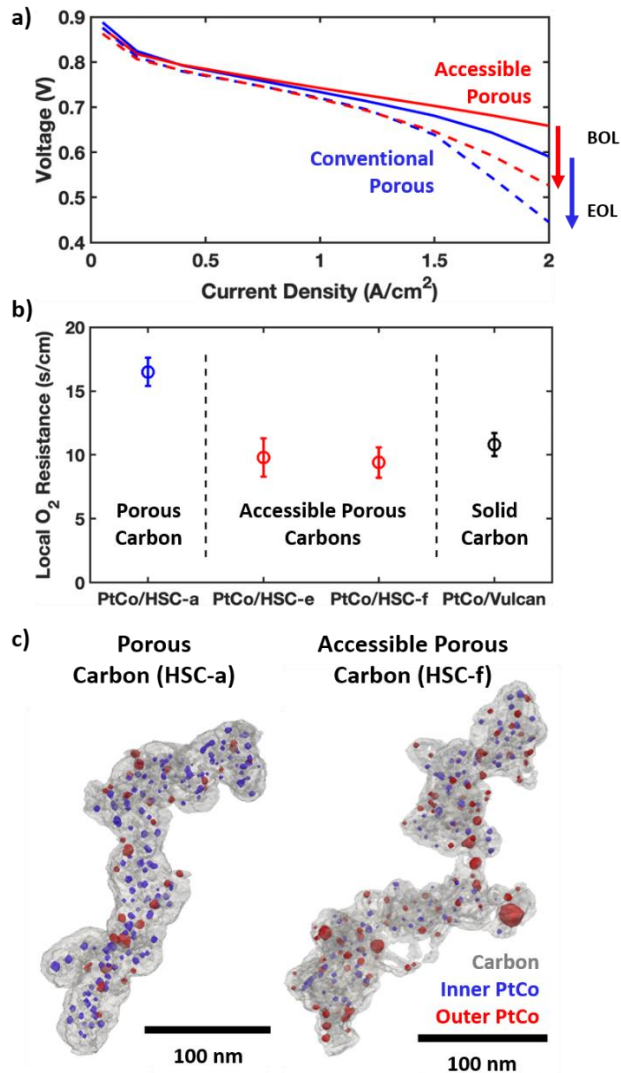

*Figure 1: (a) Polarization curves comparing the performance fuel cells using of PtCo/HSC-a (blue) and PtCo/HSC-f (red) as cathode catalysts both at beginning of life (BOL) before an accelerated stress test (solid) and at end of life (EOL) afterward (dashed). (b) Comparison of the local oxygen transport resistance of PtCo supported on standard porous carbon (HSC-a), accessible porous carbons (HSC-e and HSC-f), and solid carbon (Vulcan). (c) Surface renderings of 3D images reconstructed by electron tomography comparing PtCo catalysts supported on conventional porous carbon (HSC-a) and accessible porous carbon (HSC-f), with carbon surfaces in grey, PtCo particles inside the carbon structure in blue, and PtCo particles outside the carbon in red.*

Newly developed "accessible" porous carbons (such as the HSC-e and HSC-f carbons) were developed to mitigate these shortfalls.[7] The accessible porous catalyst supports demonstrate excellent fuel cell performance, as shown in Figure 1(a), matching the performance of standard porous supports for low-power operation, and exceeding it for high-power operation. The accessible porous catalyst supports maintain this advantage and show less degradation in high-power performance after accelerated stress testing, offering the potential to improve fuel cell durability as well. The accessible porous carbons were developed to increase the prevalence of 4-7 nm mesopores and reduce the microporosity, with the hypothesis that this would provide larger, more direct pathways for oxygen diffusion while preventing poisoning of interior Pt surfaces from intruding ionomer.[7,15] Analysis using oxygen limiting current measurements confirmed that catalysts with accessible porous supports provide lower local oxygen transport resistance (Figure 1b) in comparison to standard porous supports, thus explaining their superior high power performance.

While the accessible porous carbon supports have been investigated using electrochemical diagnostics[7,15], a detailed description of the actual physical pore structure has yet to be constructed. The overall structure of the accessible porous carbons is known to be similar to the standard Ketjenblack (referred to as HSC-a in this article), as illustrated in Figure 1(c) with surface renderings of 3D images of PtCo/HSC-a and PtCo/HSC-f created using electron tomography. Still, an understanding of the nanoscale structure of the carbons is necessary to explain differences in transport and overall fuel cell performance, and to identify potential directions for future materials development. In this study, we analyze samples of PtCo nanoparticle catalysts supported on different types of porous high-surface-area carbon (HSC), including two accessible porous samples, HSC-e and HSC-f, and the standard HSC-a (Ketjenblack) for comparison. We use transmission electron microscopy techniques, including electron tomography, to investigate the physical structure of the carbon support, its internal pores, and the distribution of PtCo particles on the carbon. In comparison to previous investigations of carbon support structure using electron tomography[11], we use low-voltage, bright-field imaging methods that minimize artifacts to provide clear visualization of the internal carbon pore structure in the presence of PtCo particles.

We then compare these microscopic structural measurements to bulk nitrogen sorption measurements of the pore structure and electrochemical diagnostics to understand the impact of the carbon support structure on reactant transport, fuel cell performance, and durability. This investigation helps clarify the role of pore size and geometry in oxygen transport in porous carbons, as well as the type of structure that leads to improved kinetics and transport in the accessible carbons. This will help direct further development of durable, high performance catalysts to enable commercial fuel cell vehicles.

## Materials and Methods

*Table 1: Segmentation measurements and electrochemistry data of HSC-a,e,f samples*

| Catalyst | Mean PtCo particle diameter (nm) | ECSA ($m^2/g_{Pt}$) | | | ORR mass activity (A/$mg_{Pt}$) | Local $O_2$ transport resistance (s/cm) |
| --- | --- | --- | --- | --- | --- | --- |
| | | Tomo. | HAD | COS | | |
| PtCo/HSC-a | 5.5 | 50.7 | 45 | 72 | 0.57 | 16.5 |
| PtCo/HSC-e | 5.4 | 45.6 | 41 | 66 | 0.65 | 9.8 |
| PtCo/HSC-f | 4.4 | 53.6 | 52 | 84 | 0.71 | 9.4 |

*Catalyst Materials*

The baseline porous carbon black is KetjenBlack EC-300J, denoted as HSC-a in this study. The carbon has a large amount of interior pores allowing Pt catalyst particles to deposit both on the inside and outside of the carbon primary particles[10,16]. This leads to transport limitation for Pt particles deposited in these interior pores. The 'accessible porous' carbons (HSC-e and HSC-f) were selected from commercial and internal sources based on their local $O_2$ transport property and ORR activity using a series of electrochemical diagnostics discussed previously[7]. PtCo nanoparticles were deposited on these three carbons to obtain a PtCo weight ratio of about 30%, followed by acid treatment to prepare dealloyed core-shell PtCo catalyst with an average Pt to Co atomic ratio of 3.[17]

Fuel cell and electrochemical evaluations of each catalyst were performed in a membrane electrode assembly with the electrocatalysts of interest on the cathode. The active area of the fuel cell single cell was 50 cm$^2$. The electrode Pt loadings were 0.1 and 0.025 mg$_{Pt}$/cm$^2$ for the cathode and anode, respectively. Perfluorosulfonic acid (PFSA) Nafion® D2020 was used in the electrode at an ionomer to carbon weight ratio of 0.8. An 18 μm thick PFSA-based membrane was used to fabricate the MEAs. A 240 μm thick carbon paper with a 30 μm thick microporous layer (MPL) coated on top was used as the gas diffusion layer (GDL). The MEAs were fabricated using a catalyst-coated-membrane approach.

*Electrochemical Measurements*

The catalyst accelerated stability test (AST), recommended by the U.S. Department of Energy (DOE), consists of 30,000 trapezoidal voltage cycles between 0.6 and 0.95 V. The dwell time at each voltage was 2.5 s and the ramp time was 0.5 s. Each cycle takes 6 s. The AST were performed at 80°C, 100% relative humidity, and ambient pressure. The fuel cell performance, electrochemically active surface area, and oxygen reduction reaction mass activity were measured at 0, 10,000 and 30,000 cycles during catalyst AST. Fuel cell polarization curve is recoded under the operating conditions in the order of anode/cathode: H$_2$/air, 94°C, 65/65%

RH$_{inlet}$, 250/250 kPa$_{abs,outlet}$, stoichiometries of 1.5/2. ORR activity is reported at 0.9 V$_{RHE}$ at 80°C, 100% relative humidity, and 1 bar of O$_2$. Pt ECSA was measured by CO stripping in an MEA.[18] For O$_2$ limiting current test, MEAs with lower Pt loading (0.06 mg$_{Pt}$/cm$^2$) and smaller active area (5 cm$^2$) were used. Current density at 0.1-0.3 V was measured at different total pressure and O$_2$ concentration at 80°C and 62% relative humidity was used to calculate the pressure independent O$_2$ transport resistance of the electrode[19,20].

*Electron Microscopy Sample Preparation*

Catalyst powders were dispersed in ethanol by ultrasonication. Several microliters of this mixture were dropped on 100-mesh hexagonal copper grids coated with an ultrathin (nominally 3-4 nm) carbon film. These grids are well suited for tomography experiments by allowing a large tilt range and low background signal. To eliminate the possibility of etching the carbon supports, the samples were not plasma cleaned.

*Transmission Electron Microscopy Imaging and Tomography*

Transmission electron microscopy (TEM) was carried out on an FEI T12 BioTwin with a lanthanum hexaboride filament. An accelerating voltage of 80kV was used to reduce beam damage to carbon in the samples while maintaining high imaging resolution and a tolerable effective sample thickness for tomography. Bright field imaging was used to provide strong phase contrast for carbon and a relatively small contrast ratio between platinum and carbon, allowing for improved visualization of the carbon structure and pores. No objective aperture was used.

Electron tomography tilt series data was acquired manually using a Fischione Model 2020 tomography holder. Images were acquired over the full range of sample visibility, typically around 150° total, at either one- or two-degree increments. Tomography was conducted at room temperature. The bright field TEM imaging mode mitigates carbon contamination concerns for tomography because contaminant carbon is typically deposited outside of the field of view before reaching the area of interest.

*Tomography Reconstruction*

Alignment and reconstruction of tomography data was done in Matlab using custom tools. Images in each tomography tilt series were aligned manually to a common coordinate system using a single PtCo particle visible at all tilt angles as a fiducial marker. Low quality images where the fiducial marker is not very visible and the signal is weak were removed. Alignments were accurate to approximately one pixel in each direction, with pixel to length ratio of 1.7 pixels per nm. Aligned tilt series background intensity was removed by dividing by the modal intensity of all pixels from the tilt series. Tomograms were reconstructed using a weighted back projection algorithm implemented using the *iradon* function in Matlab.

Segmentation

Reconstructed tomograms of each of the samples was segmented to identify PtCo particles using a threshold and active contour that employs the Chan-Vese method[21]. Artifacts in the PtCo segmentation are smoothed using morphological open and close operations and nearby particles that are erroneously segmented as a single particle are separated using a watershed transform. Statistics on PtCo particle size and geometry are calculated using Matlab's *regionprops3* function. The phase contrast mechanism of bright field TEM makes the pore structure of the carbon support more easily visualized, but relatively difficult to segment automatically. As such this study did not use a carbon segmentation for quantitative measurements. PtCo particles on the exterior and interior of the carbon support were identified using a custom semi-manual sorting GUI, which streamlined human identification of particle locations.

3D electron tomography visualizations were created using the open source tomography platform tomviz (http://www.tomviz.org/)[22]. Visualizations of segmented reconstructions were created using binary volumes for each material component exported from Matlab created with medium thresholds. Surfaces are contour renderings at the value 0.5 for each binary component.

*Quantitative Measurements of 2D and 3D TEM Images*

Carbon primary particle diameter were measured manually from sets of 2D TEM images using ImageJ including a range of carbon particles in each sample. Carbon pore sizes were measured manually from cross-sections of reconstructed tomograms using ImageJ. Measurements of segmented PtCo particles such as equivalent diameter were calculated using the *regionprops3* function in Matlab and surface area was calculated using the equivalent diameter and spherical approximation (See supplemental Figure 3). Measurements were made separately for each tomography dataset and final measurements report combined data for HSC-a and HSC-e. See the supplemental information for a more in-depth discussion and measurements reported separately for each tomography dataset.

# Results & Discussion

*Overview of Catalyst, Support, and Pore structure*

Three different PtCo catalysts were examined for this study: one deposited on a conventional porous carbon, KetjenBlack EC-300J (PtCo/HSC-a) and two with 'accessible' porous carbon supports (PtCo/HSC-e and PtCo/HSC-f). HSC-e was modified from KetjenBlack, while HSC-f is a similar carbon but with smaller carbon primary particles[7]. As such, both accessible carbons were aimed to facilitate $O_2$ transport close to the Pt particles by either enlarging the carbon pore size or shortening the transport length of $O_2$. Catalysts on both carbons have shown superior fuel cell performance, particularly at high current density where local $O_2$ transport becomes the limiting factor[7].

The morphology of the carbon support was first examined using TEM images (Figure S1). Within each sample, the images indicate a variation in the density of the carbon support as reported previously[5], where some areas of the carbon appear to have relatively hollow interiors and other areas of the carbon relatively filled interiors. This variation of carbon morphology appears broader, possibly bimodal, in HSC-a and HSC-e. In contrast, HSC-f is relatively uniform, with the carbon generally appearing hollow. Carbon primary particle diameter appears to be the same in HSC-a and HSC-e, but slightly smaller in HSC-f. There is no observable difference between the size of primary particles of hollow and full carbons. Representative TEM images of these different carbon types are shown in Figure 2.

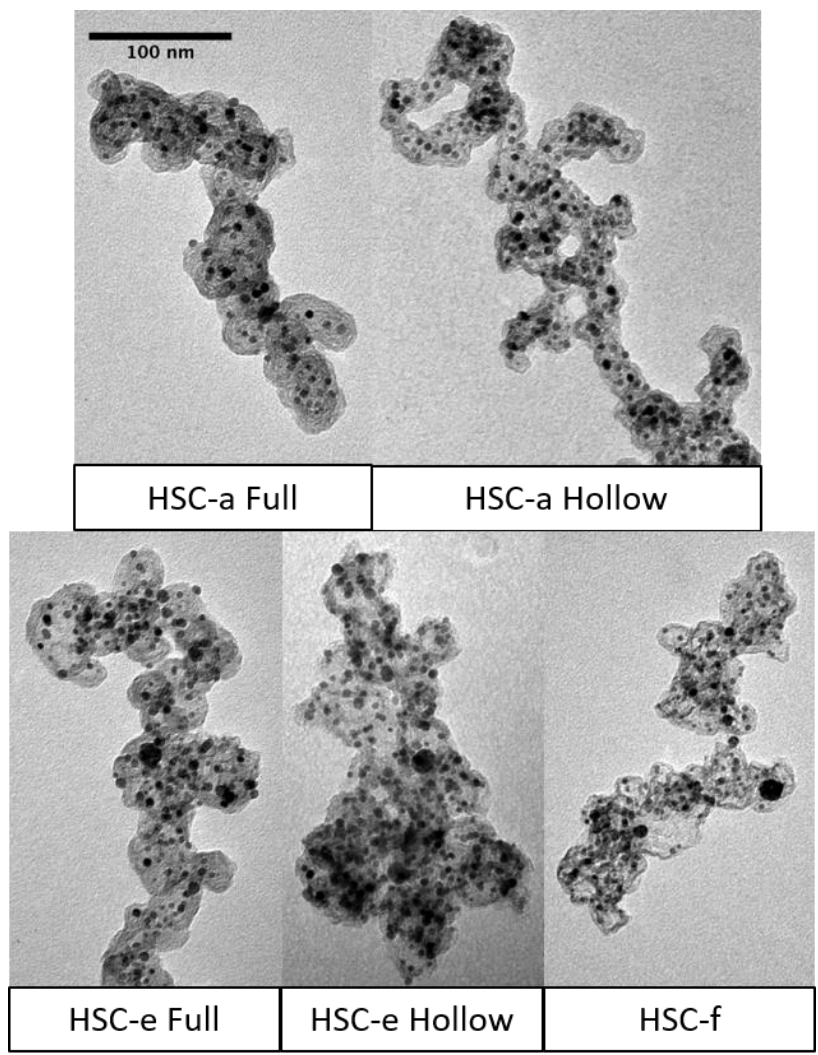

*Figure 2: Representative TEM images for calalysts supported on conventional (HSC-a) and accessible (HSC-e, HSC-f) porous carbons. Typical full and hollow particles of HSC-a and HSC-e were chosen to represent morphology variation within those samples. All images are at the same scale.*

To understand the details of the internal pore structure of these different carbons, we use TEM tomography to reconstruct 3-dimensional images of each catalyst sample. To capture the variation of carbon structure in the HSC-a and HSC-e samples, two tomography reconstructions were produced for each, from particles representative of the relatively hollow and relatively filled regions of the carbon. Tomography was carried out on the 5 representative particles shown in Figure 2.

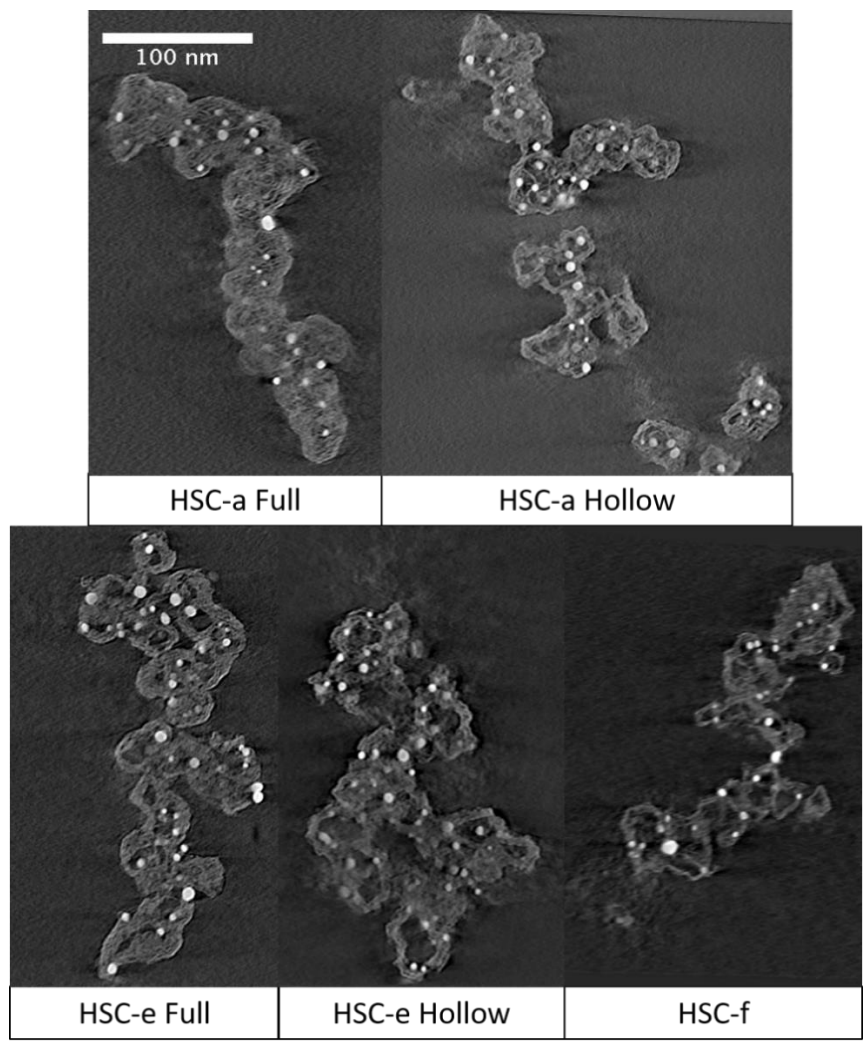

*Figure 3: Representative tomography cross sections for each particle from Figure 2 showing an overview of pore structure. Porosity transitions from predominantly microporous volume in HSC-a to predominantly mesoporous volume in HSC-f. Contrast from bright-field TEM is inverted so that the background is dark, PtCo particles are bright, and carbon is intermediate intensity. All cross sections are at the same scale and perpendicular to the imaging axis.*

The reconstructed tomograms are visualized in cross-section in Figure 3 to allow comparison of the particles' internal pore structure. In HSC-a, the relatively full particles are predominantly microporous, with the largest pores similar in size to the PtCo particles. The more hollow particles in HSC-a and the accessible carbons have larger mesopores in the carbon interior. The large internal mesopores in the accessible carbons result in relatively thin carbon shells surrounding them, especially in HSC-f, which has smaller primary particles, but similarly sized mesopores compared to HSC-e.

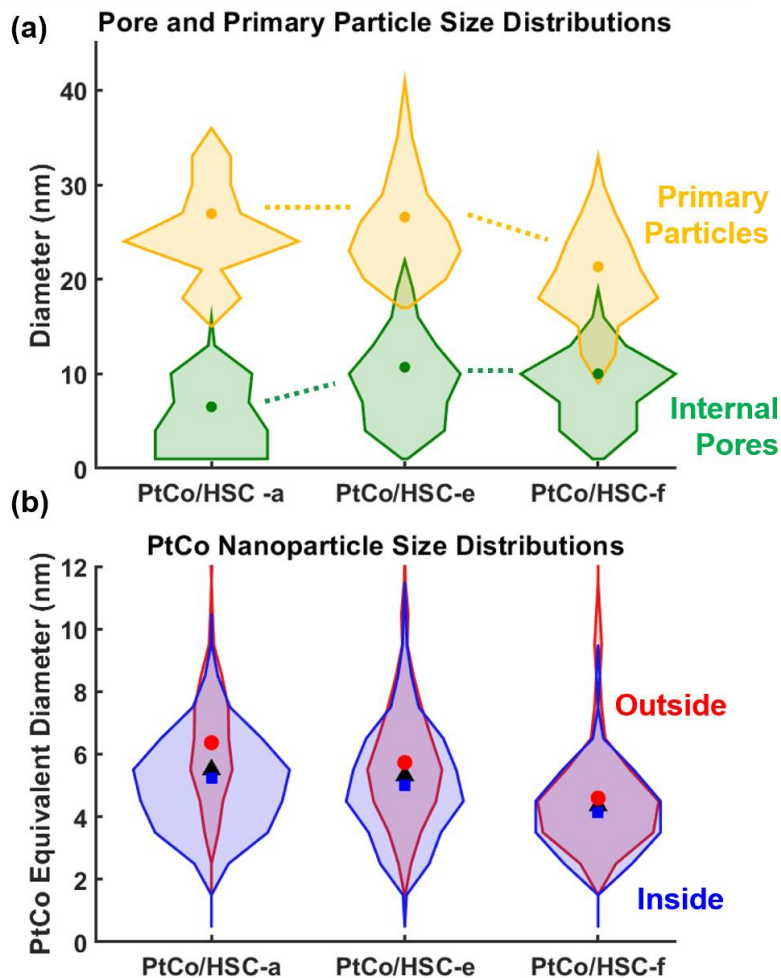

*Figure 4: Size distributions from electron tomography of (a) internal pore diameters (green) and carbon primary particle diameters (yellow) and (b) PtCo Nanoparticle diameter for particles inside (blue) and outside (red) the carbon support. Distributions are histograms of pore and particle counts, with the mean diameter of the distribution marked by points. The black point is the overall mean of all particles.*

Measurements of the carbon primary particle and internal mesopore sizes are shown in Figure 4(a). In HSC-a, primary particles average approximately 26nm in diameter, while internal mesopores average around 7 nm and range from around 3 to 10 nanometers (20$^{th}$-80$^{th}$ percentiles). In HSC-e, the size of primary particles is unchanged relative to HSC-a, but the size of internal mesopores has grown significantly, to an average of around 11 nm with a range from around 7 to 14 nanometers. HSC-f has relatively small primary particles, averaging around 21 nm, but similarly sized internal mesopores compared to HSC-e, with an average of around 10 nm with a range from around 7 to 13 nanometers. The internal mesopores in the accessible carbons are thus significantly larger than the 2-8 nm range assumed by previous investigations[7,15] in their interpretation of nitrogen sorption measurements. Furthermore, the thickness of the carbon shell separating interior mesopores from the support exterior, which depends on both the primary particle sizes and mesopore sizes, decreases from HSC-a to HSC-e to HSC-f.

The statistical distribution of PtCo particles on the inside and outside of the carbon support (Figure 4(b)) for each sample was measured to determine the impact of the carbon pore and primary particle structure. In HSC-a, approximately 75% of the PtCo particles are found in the carbon interior. The fraction of nanoparticles on the interior of the carbon decreases with the observed increase in mesoporosity, with around 60% on the interior for HSC-e and around 50% on the interior for HSC-f. This is because during Pt deposition, the number of nucleation sites closely correlates with the carbon surface area. Therefore, higher interior surface area (more micropores) will result in a larger fraction of interior Pt.

In HSC-a, as expected for porous carbon catalysts, interior PtCo particles are restricted by the small internal pores and are thus smaller than exterior PtCo particles. . Interestingly, this difference in the size of interior and exterior particles is less noticeable for HSC-e and HSC-f. We also compared the interior and exterior PtCo size distributions for regions of the HSC-a and HSC-e samples with relatively hollow vs full carbons, and find similar trends (Figure S2).

From the tomography, we then calculated the specific surface area, and found it is very similar (within ~10%) to the electrochemically-active surface area (ECSA) measured by hydrogen adsorption-desorption (HAD) (Table 1). However, the ECSA measured by CO-stripping is consistently about 50% higher than that measured by either other method.

*Diffusion Pathways into the Support Interior*

In our earlier study[11], narrow (~1 nm) channels were observed connecting the internal carbon mesopores to the support exterior in HSC-a. This led us to hypothesize that narrow pore openings restrict $O_2$ transport to the interior Pt, and that enlarging these opening could improve fuel cell performance. If present, larger pore openings could explain the improved high-current-density performance of catalysts supported on accessible porous carbons.

However, mesopore-sized openings (>2 nm width) are rarely observed in the tomography experiments performed here for HSC-a, HSC-e, and HSC-f, and are too infrequent to quantify for any of the carbons. Although HSC-f may have slightly larger and more visible pore openings, these openings do not seem abundant enough to be the primary pathway for gas diffusion into the carbon interior.

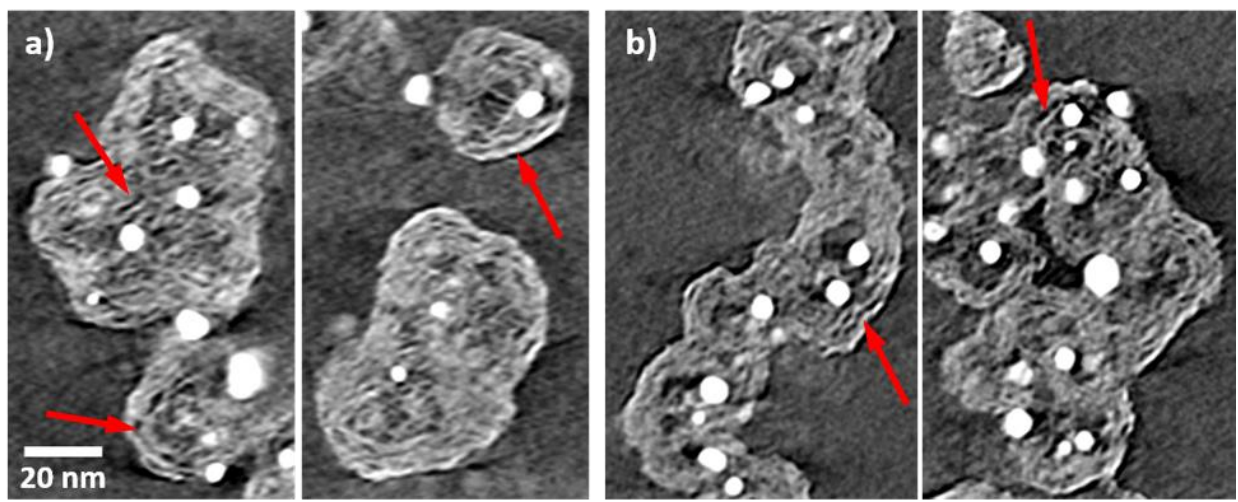

*Figure 5: Closeup of representative tomography cross sections for HSC-a (a) and HSC-e (b) showing the microporous structure in solid carbon. Contrast from bright-field TEM is inverted so that the background is dark, PtCo particles are bright, and the carbon support is intermediate intensity. All cross sections are the same scale and perpendicular to the imaging axis.*

Instead, tomographic imaging (Figure 5) suggests that the carbon micropores (<2 nm width) may be an important pathway for oxygen transport into the support interior. In all the carbons, ~1-2 nm thick micropores are present permeating through the relatively solid carbon regions and the shells surrounding mesopores. These micropores appear to commonly have a slit-like shape, with narrow ~1-2 nm widths and a wide distribution of lengths, commonly ranging from 5-10 nm. Slit-shaped micropores are well known to be present in carbon materials[23,24], resulting from gaps in between semi-graphitic layers of carbon. The slit-shaped micropores visible in Figure 5 appear to commonly form an onion-like concentric structure in the round primary carbon particles. This type of slit-shaped micropore may be especially prevalent in HSC-type carbons because of the higher degree of disorder in the carbon relative to more graphitic carbons such as Vulcan or graphitized carbon blacks. Even in the absence of larger, mesopore-like openings, the interior of the carbon supports may remain accessible through microporous channels due to the large number of these channels along with their relatively high cross sectional area.

The accessibility of interior pores was investigated further by comparison to nitrogen adsorption and desorption measurements (BET-BJH), as shown in Figure 6. The nitrogen adsorption and desorption in Figure 6(a) show a characteristic of Type 4 isotherms and combined Type H3 and H4 hysteresis loops according to IUPAC classification[25], as expected for powder materials with a large degree of mesoporosity and a wide pore size distribution. The hysteresis loops from 0.45-0.85 relative pressure appear to get wider for the accessible carbons, especially

HSC-f, in comparison to HSC-e.  This is likely a result of the greater mesoporosity in the accessible carbons and also suggests pore blocking effects, where the mesoporous volume is restricted by small pore openings.[23,25]  Overall, there is significantly more nitrogen adsorbed on HSC-f than on either HSC-a or HSC-a, most likely because of the smaller size of the carbon primary particles in HSC-f.

Pore size analysis from these nitrogen sorption isotherms (Figure 6b) is complicated by the tensile strength effect (TSE).[23,26]  The TSE causes isotherm hysteresis loops to close at a critical relative pressure around 0.45, below which the hemispherical meniscus that typically forms during desorption becomes unstable and pores empty abruptly by cavitation. This leads to the particularly large peaks at ~3.8 nm in the desorption isotherms in Figure 6(b). It is important to note that this peak does not correspond to a real feature of the sample pore structure, and that pores with openings smaller than the 3.8 nm critical diameter will empty at this point as well. Because of both the TSE and pore blocking effects, use of the desorption isotherm may lead to significant inaccuracies in calculated pore size distributions.

The adsorption isotherm is mostly unaffected by the TSE and pore blocking effects, and therefore more accurately reflects the size distribution of the internal pores. In the pore size distribution calculated from the adsorption isotherm it is apparent that there is significantly more mesopore volume across the 3-16 nm size range in HSC-e and HSC-f when compared to HSC-a. This increase is even more pronounced for HSC-f because of the smaller carbon primary particles, resulting in comparatively more mesopore volume per mass. This result is consistent with the differences in carbon pore structure and primary particle size observed through TEM tomography (Figure 3 and Figure 4a).  It is also notable that in the desorption size distributions the expected increase in mesopore volume is absent, but the TSE peak is much larger for the accessible porous carbons. This is expected based on the structure observed through TEM tomography, where the internal mesopores surrounded by microporous carbon and are connected to the carbon exterior through micropores smaller than the TSE critical size.

Oxygen diffusing to PtCo particles in the carbon support interior therefore likely passes predominantly through these microporous channels. Despite their apparent tortuosity, these micropores may be capable of supporting significant oxygen diffusion because of the large open volume and cross-sectional area created by their slit-shaped geometry. The increased mesoporosity in HSC-e and HSC-f and decreased primary particle size in HSC-f, lead to a shorter pathlength for diffusion through the micropores in the carbon shell. This shorter pathlength is likely responsible for the decreased local oxygen transport resistance in the accessible porous carbons HSC-e and HSC-f.

It is also possible that empty pore volume in the immediate vicinity of PtCo particles on the carbon interior is an important factor in effective local oxygen transport. In HSC-a, pores are often only slightly larger than PtCo particles contained in them, which may limit the rate of oxygen diffusion to the surface. The larger mesopores in HSC-e and HSC-f create some open space surrounding embedded PtCo particles, where the oxygen flux is highly concentrated.

These results explain why the accessible carbons simultaneously achieve good local $O_2$ transport and high ORR kinetic activity. The larger interior pores and shorter, less tortuous diffusion paths together help lower the $O_2$ transport resistance. At the same time, the small microporous openings in the carbon shell effectively exclude ionomer from entering carbon pores and poisoning the Pt surfaces, enabling the high ORR mass activities (Table 1).

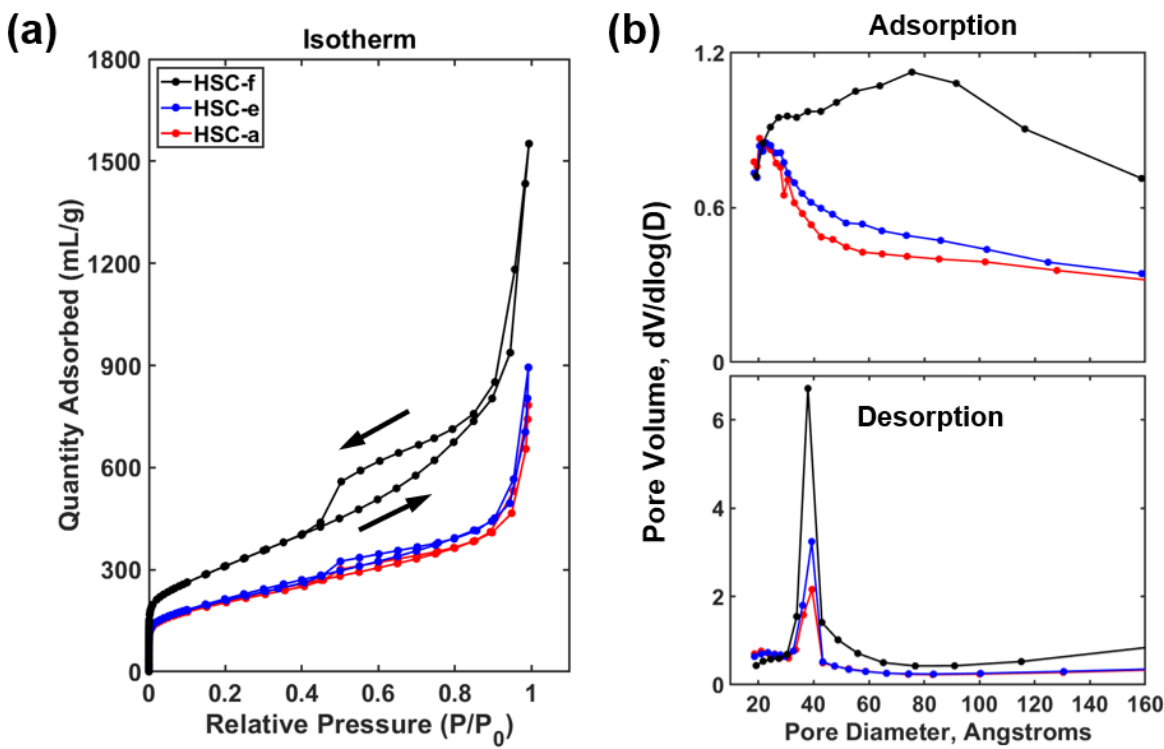

*Figure 6: Nitrogen adsorption-desorption analysis of PtCo catalysts supported on HSC-a, HSC-e, and HSC-f.(a) Nitrogen adsorption desorption isotherms and (b) pore size distributions calculated using the BJH technique for both the adsorption (top) and desorption (bottom) isotherms.*

<u>*Consequences of Support Pore Structure for Catalyst Durability*</u>

Figure 7 shows a comparison of the durability of PtCo catalysts supported on the three porous carbons and Vulcan carbon subjected to a catalyst-targeting accelerated stress test (AST). The AST involves potential cycling between 0.6 and 0.95 V vs RHE at 80°C and 100% RH. It is designed to accelerate Pt coarsening via Pt dissolution and redeposition without causing significant oxidative degradation of the carbon support. At beginning of life, all three of the catalysts with porous carbon supports have higher ECSA (Figure 7(a)) and mass activity (Figure 7(b)) compared to PtCo/Vulcan, and the catalysts with accessible porous supports have an advantage in high current density performance (Figure 7(c)). However, the accessible porous carbons, particularly HSC-f, have significantly greater ECSA loss after the AST (Figure 7(a)) in comparison to HSC-a. This is likely caused by higher rates of coalescence for internal particles in mesopores, which are not as confined as particles in the small pores of HSC-a. There may also be some contribution from the relatively small PtCo particle size in HSC-f, which makes it more vulnerable to electrochemical Ostwald ripening. The more rapid ECSA loss for the accessible porous catalysts also leads to a greater loss of mass activity (Figure 7(b)), although the magnitude of this loss is somewhat less due to the beneficial effect of larger particle sizes on the specific activity.

Despite these large ECSA losses, surprisingly, the accessible porous carbons still retain the greater HCD performance after the AST (Figure 7(c)). Typically, large ECSA losses lead to serious losses in high current density performance due to ineffective local oxygen transport[1,27–29]. However, the reduction in local oxygen transport resistance provided by the accessible porous carbon structure dramatically mitigates the negative impact of ECSA loss on HCD performance. This effect more than compensates for the more rapid ECSA loss caused by the more open support interior.

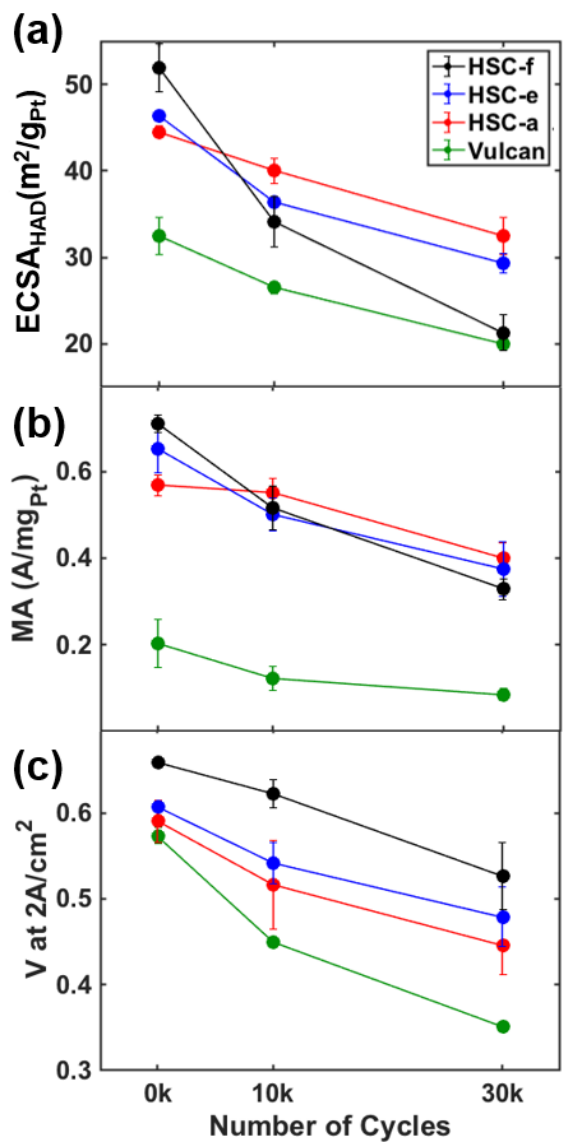

*Figure 7: Durability during accelerated stability test of PtCo catalysts supported on accessible porous carbons (HSC-e, HSC-f), conventional porous carbon (HSC-a) and Vulcan carbon for (a) $ECSA_{HAD}$, (b) ORR mass activity, and (c) Voltage at 2 A/cm$^2$.*

## Conclusion

New accessible porous carbon supports (HSC-e and HSC-f), which were previously shown to provide both good kinetics and high current density performance, were studied using TEM imaging and tomography and compared to a conventional porous carbon (HSC-a) to determine what characteristics of the carbon morphology contributed to their improved performance. Cross-sections of tomographic reconstructions show that all three porous carbons contain internal mesopores as well as narrow, ~1-2 nm wide microporous channels, which appear to have a flat, slit-like geometry formed between carbon sheets. Mesopore-sized (~3-4 nm or larger) pore openings appear rare and are therefore unlikely to support significant oxygen transport to the interior. We propose that microporous channels connecting interior mesopores and the carbon exterior are the primary pathway for oxygen transport to the support interior. This conclusion is supported by nitrogen sorption isotherms and BJH pore size distributions, which indicate that nitrogen adsorbed in mesopores desorbs through smaller, microporous openings. The slit-shaped geometry of the micropores may enable significant oxygen diffusion by creating a relatively large open volume and cross-sectional area. The lack of large pore openings also ensures that ionomer infiltration into the support interior will be blocked to preserve highly active catalyst surfaces.

The accessible porous carbons HSC-e and HSC-f have internal mesopores that are relatively large (ranging from 7 to 14 nm) and prevalent in comparison to the mesopores in HSC-a (ranging from 3 to 10 nm). HSC-f also has somewhat smaller carbon primary particles, averaging around 21 nm, compared to around 26 nm for both HSC-a and HSC-e, leading to a thinner shell of carbon surrounding the internal mesopores. The structure of the accessible porous carbons may lead to improved local oxygen transport in two ways. First, the thinner carbon shell surrounding interior mesopores and PtCo particles allows for a shorter, less tortuous path for oxygen diffusion through the microporous channels. Second, the larger mesopores in the accessible carbons also prevent PtCo particles from being closely confined by carbon, allowing for higher oxygen flux in the immediate vicinity of the Pt surface. Further investigations with transport modelling may be able to distinguish the relative importance of these two factors.

The accessible carbons also tend to have a higher fraction of PtCo particles and active surface on the carbon exterior. This is likely because the larger interior pores lead to less interior surface area and thus fewer interior nucleation sites for PtCo particles. Development of accessible porous carbons with more internal surface area may thus allow further improvements in catalyst kinetics. The larger mesopores and higher fraction of exterior PtCo particles for the accessible porous carbons may also contribute to contribute to PtCo particle coalescence during accelerated stress testing, leading to the more severe ECSA and mass activity loss compared to HSC-a. Despite this, the accessible porous carbons maintain higher performance at high current densities because the improved local transport resistance decreases the performance penalty of ECSA losses. Accessible porous supports therefore offer a potential advantage for fuel cell durability as well as for performance. It is possible that accessible porous carbons could be further improved to extend durability through other approaches to suppress coalescence, such as anchoring of the catalyst particles.

# Acknowledgements

This work was supported by the U.S. Department of Energy, Office of Energy Efficiency and Renewable Energy under grant DE-EE0007271 (pre-2019), and through the Center for Alkaline-based Energy Solutions, an Energy Frontier Research Center funded by the US Department of Energy, Office of Science, Basic Energy Sciences Award DE-SC0019445 (2019 and later). Matthew Ko acknowledges support from the Rawlings Cornell Presidential Research Scholarship. Elliot Padgett acknowledges support from an NSF Graduate Research Fellowship (DGE-1650441). This work made use of electron microscopy facilities at the Cornell Center for Materials Research supported by the NSF MRSEC program (DMR-1719875). The authors thank Dr. Mariena Silvestry Ramos and John Grazul for assistance with electron microscopy facilities and sample preparation, as well as Michael Cao for assistance with microscope operation. Dr. Barnaby Levin performed preliminary TEM experiments.

# References

1. A. Kongkanand and M. F. Mathias, *J. Phys. Chem. Lett.*, **7**, 1127–1137 (2016).

2. H. A. Gasteiger, S. S. Kocha, B. Sompalli, and F. T. Wagner, *Appl. Catal. B Environ.*, **56**, 9–35 (2005).

3. N. Konno, S. Mizuno, H. Nakaji, and Y. Ishikawa, *SAE Int. J. Altern. Powertrains*, **4**, 123–129 (2015).

4. X. Tuaev, S. Rudi, and P. Strasser, *Catal. Sci. Technol.*, **6**, 8276–8288 (2016) http://xlink.rsc.org/?DOI=C6CY01679K.

5. E. Padgett et al., *J. Electrochem. Soc.*, **166**, F198–F207 (2019) http://jes.ecsdl.org/lookup/doi/10.1149/2.0371904jes.

6. Y.-C. Park, H. Tokiwa, K. Kakinuma, M. Watanabe, and M. Uchida, *J. Power Sources*, **315**, 179–191 (2016) http://dx.doi.org/10.1016/j.jpowsour.2016.02.091.

7. V. Yarlagadda et al., *ACS Energy Lett.*, **3**, 618–621 (2018)

https://pubs.acs.org/doi/10.1021/acsenergylett.8b00186.

8. K. Kodama et al., *J. Electrochem. Soc.*, **161**, 649–652 (2014).

9. K. Shinozaki, Y. Morimoto, B. S. Pivovar, and S. S. Kocha, *J. Power Sources*, **325**, 745–751 (2016) http://dx.doi.org/10.1016/j.jpowsour.2016.06.062.

10. I. Toshihiro, *Electrochem. commun.*, **31**, 412–421.

11. E. Padgett et al., *J. Electrochem. Soc.*, **165**, F173–F180 (2018) http://jes.ecsdl.org/lookup/doi/10.1149/2.0541803jes.

12. A. Kongkanand, V. Yarlagadda, T. Garrick, T. E. Moylan, and W. Gu, *ECS Trans.*, **75**, 25–34 (2016).

13. T. R. Garrick, T. E. Moylan, V. Yarlagadda, and A. Kongkanand, *J. Electrochem. Soc.*, **164**, F60–F64 (2017) http://jes.ecsdl.org/lookup/doi/10.1149/2.0551702jes.

14. N. Ramaswamy and S. Kumaraguru, *ECS Trans.*, **85**, 1449–1457 (2018).

15. N. Ramaswamy, W. Gu, J. M. Ziegelbauer, and S. Kumaraguru, *J. Electrochem. Soc.*, **167**, 064515 (2020).

16. H. Jinnai, R. J. Spontak, and T. Nishi, *Macromolecules*, **43**, 1675–1688 (2010).

17. B. Han et al., *Energy Environ. Sci.*, **8**, 258–266 (2015).

18. T. R. Garrick, T. E. Moylan, M. K. Carpenter, and A. Kongkanand, *J. Electrochem. Soc.*, **164**, F55–F59 (2017).

19. D. R. Baker, D. A. Caulk, K. C. Neyerlin, and M. W. Murphy, *J. Electrochem. Soc.*, **156**, B991 (2009).

20. T. A. Greszler, D. Caulk, and P. Sinha, *J. Electrochem. Soc.*, **159**, F831–F840 (2012) https://iopscience.iop.org/article/10.1149/2.061212jes.

21. T. F. Chan and L. A. Vese, *IEEE Trans. Image Process.*, **10**, 266–277 (2001) https://www.math.ucla.edu/~lvese/PAPERS/IEEEIP2001.pdf.

22. B. D. A. Levin et al., *Micros. Today*, **26**, 12–17 (2018).

23. T. Horikawa, D. D. Do, and D. Nicholson, *Adv. Colloid Interface Sci.*, **169**, 40–58 (2011).

24. H. Kurig et al., *Carbon N. Y.*, **100**, 617–624 (2016) http://dx.doi.org/10.1016/j.carbon.2016.01.061.

25. M. Thommes et al., *Pure Appl. Chem.*, **87**, 1051–1069 (2015).

26. J. C. Groen, L. A. A. Peffer, and J. Pérez-Ramírez, *Microporous Mesoporous Mater.*, **60**, 1–17 (2003).

27. R. K. Ahluwalia et al., *J. Electrochem. Soc.*, **165**, F3316–F3327 (2018).

28. A. Kneer et al., *J. Electrochem. Soc.*, **165**, F3241–F3250 (2018).

29. G. S. Harzer, J. N. Schwämmlein, A. M. Damjanović, S. Ghosh, and H. A. Gasteiger, *J. Electrochem. Soc.*, **165**, F3118–F3131 (2018).